\documentclass[10pt,aps,prd,preprintnumbers,superscriptaddress,twocolumn]{revtex4-2}
\usepackage[dvips]{graphicx}
\usepackage{bm,latexsym,amsmath,amssymb,amsfonts,mathrsfs}
\usepackage{mathtools}
\usepackage[colorlinks=true,urlcolor=blue,anchorcolor=black,citecolor=blue,linkcolor=black,filecolor=black,menucolor=black,linktocpage=true,pdfproducer=medialab,pdfa=true]{hyperref}
\usepackage{xcolor}
\usepackage{comment}

\begin{document}
\preprint{TTI-MATHPHYS-41}

\vspace*{0.5cm}

\title{Trapped Surface as a Cosmic Censor}

\author{Hideo Furugori}
\email{hideo@toyota-ti.ac.jp}
\affiliation{Mathematical Physics Laboratory, Toyota Technological Institute, Nagoya 468-8511, Japan}

\author{Daisuke Yoshida}
\email{yoshida.daisuke.k9@f.mail.nagoya-u.ac.jp}
\affiliation{
Institute for Advanced Research, Nagoya University, Nagoya 464-8602, Japan}

\author{Kaho Yoshimura}
\email{yoshimura-kaho848@g.ecc.u-tokyo.ac.jp}
\affiliation{Graduate School of Arts and Sciences, University of Tokyo, Komaba,
Tokyo 153-8902, Japan}

\begin{abstract}
We formulate a local geometric criterion for weak cosmic censorship in black hole overcharging and overspinning thought experiments. Under the null convergence and generic conditions, matter injection turns a horizon cross section into a closed trapped surface. Any final spacetime unable to accommodate this surface is ruled out. This trapped surface criterion excludes superextremal Reissner-Nordstr\"om, Reissner-Nordstr\"om-de Sitter, and Kerr-Newman final states, as well as Weyl-class naked singularities. Our criterion does not rely on asymptotic charges or on an extremal condition characterizing naked singularities.
\end{abstract}

\maketitle

\section{Introduction}
Whether spacetime singularities can be visible to distant observers is a fundamental question in general relativity. The weak cosmic censorship conjecture asserts that singularities produced by gravitational collapse are generically hidden behind event horizons \cite{Penrose:1969pc}. Although this conjecture has been examined in a wide variety of black hole spacetimes,
it remains an open question how generally one can characterize the conditions that prevent the formation of naked singularities.

A standard way to test this conjecture is to consider overcharging/overspinning thought experiments in which energy, angular momentum, and charge are injected into a black hole from the outside~\cite{Wald:1974hkz}. 
While extremal black holes cannot be overcharged or overspun in the test particle approximation~\cite{Wald:1974hkz, Cohen:1979zzb, Needham:1980fb, hiscock1981magnetic, Gwak:2015fsa, Duztas:2016xfg, Ghosh:2019dzq, Yang:2020iat, Izumi:2024rge}, slightly subextremal black holes may appear vulnerable within this approximation~\cite{Hubeny:1998ga, deFelice:2001wj, Hod:2002pm, Jacobson:2009kt, Chirco:2010rq, Saa:2011wq, Gao:2012ca, Duztas:2016xfg, Yang:2020iat}. 
These studies have revealed the importance of backreaction in assessing whether weak cosmic censorship can be violated.
The perturbative analysis developed by Sorce and Wald~\cite{Sorce:2017dst}, based on the covariant phase space formalism, rules out the overcharging and overspinning of slightly subextremal Kerr-Newman black holes by matter satisfying the null energy condition.
The Sorce-Wald formalism has become a standard framework for testing weak cosmic censorship beyond the test-particle approximation, and has been applied to a wide range of black hole spacetimes
~\cite{An:2017phb,Ge:2017vun,Duztas:2018fbc,Chen:2019nhv,Shaymatov:2019del,Shaymatov:2019pmn,Jiang:2019ige,Jiang:2019vww,Wang:2019bml,He:2019mqy,Jiang:2019soz,Jiang:2020btc,Jiang:2020mws,Wang:2020vpn,Shaymatov:2020byu,Jiang:2020alh,Qu:2020nac,Jiang:2020xow,Zhang:2020txy,Li:2020smq,Ding:2020zgg,Duztas:2021kni,Qu:2021hxh,Sang:2021xqj,Huang:2022avq,Sang:2022lfk,Ahmed:2022dpu,Wang:2022umx,Jiang:2022zod,Yang:2023hll,Duztas:2024vky,Yoshida:2024txh}. 
In its standard implementation, however, the fluxes of energy, angular momentum, and charge through the horizon are related to asymptotically defined conserved charges. Then, the formalism allows us to test whether the resulting parameters violate the extremal bound within a prescribed family of final spacetimes.

Such a formulation in terms of asymptotic charges
is naturally applied in asymptotically flat or asymptotically anti-de Sitter spacetimes, but is less direct in de Sitter spacetimes. 
The prevention of overcharging in de Sitter settings~\cite{Yoshida:2024txh} 
suggests that the underlying censorship mechanism may 
be formulated without reference to asymptotically defined charges.
Such a formulation would also be useful in spacetimes with nonstandard asymptotics, where global charges can be subtle.

In this paper, we identify a local geometric mechanism behind the weak cosmic censorship in black hole overcharging/overspinning thought experiments.
The key observation is that, under the null convergence condition,
which corresponds to the null energy condition through the Einstein equations, the injected matter generically focuses the null generators of a Killing horizon of the initial black hole, turning a horizon cross section into a closed trapped surface.
This leads to our trapped surface criterion: \emph{any candidate final state incompatible with the existence of this trapped surface is excluded.}
Since the trapped surface formation is based on the perturbed Einstein equations, this criterion incorporates the backreaction of the injected matter. We formulate this criterion in Sec.~\ref{sec: setup}.

At first sight, the formation of a trapped surface may appear to settle the issue immediately, since trapped surfaces are commonly associated with black hole formation. The Penrose singularity theorem~\cite{Penrose:1964wq} relates the existence of a trapped surface to singularity formation, but does not by itself determine whether the resulting singularity is naked. A rigorous statement relating trapped surfaces to the black hole region is the Hawking-Ellis theorem~\cite{Hawking:1973uf,Wald:1984rg,Claudel:2000wb}, which places trapped surfaces inside the black hole region only under assumptions that already exclude naked singularities. 
Neither theorem therefore provides a suitable criterion for testing cosmic censorship.
Instead, one must directly examine whether the candidate final spacetime can accommodate the trapped surface.

We demonstrate how trapped surface formation acts as a cosmic censor in Reissner-Nordstr\"om, Reissner-Nordstr\"om-de Sitter, and Kerr-Newman spacetimes.
These demonstrations represent three geometrically distinct ways in which superextremality is related to the absence of trapped surfaces. 
The Reissner-Nordstr\"om case is the simplest: the static Killing field is timelike in the entire region for superextremal case, and Mars-Senovilla's theorem~\cite{Mars:2003ud} forbids a closed trapped surface entirely contained in such a region.
In the Reissner-Nordstr\"om-de Sitter and Kerr-Newman cases, although additional local checks are required, we show in Sec.~\ref{sec: case studies} that the corresponding superextremal final states cannot accommodate the trapped surface. Therefore, the superextremal final states are excluded in these examples, recovering the conclusions of Sorce-Wald~\cite{Sorce:2017dst}. In addition, because the argument for the Reissner-Nordstr\"om case relies only on the existence of a region with a timelike Killing vector that contains the relevant surface, it applies to a broader class of candidate final spacetimes.

\section{Trapped surface criterion}
\label{sec: setup}
\begin{figure}[t]
 \centering
 \includegraphics[width=0.45\textwidth]{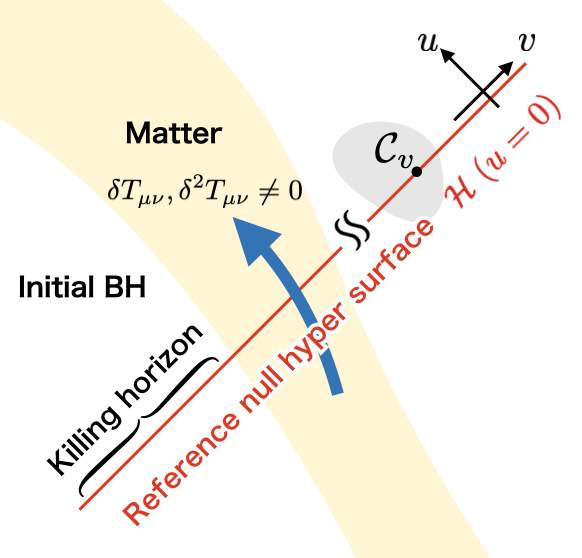}
 \caption{The setup of the overcharging/overspinning thought experiment. $\mathcal{C}_{v}$ is the surface shown to become trapped. Our criterion can constrain a final state based on the local geometry around $\mathcal{C}_{v}$ (grey region). }
 \label{fig:setup}
\end{figure}

We consider a process in which matter is injected into an initially stationary black hole during a finite time interval as shown in Fig.~\ref{fig:setup}.
We ask which candidate final spacetime can consistently describe the geometry after the injection.
We describe this process perturbatively around the initial black hole spacetime.

We now formulate this setup rigorously \cite{Sorce:2017dst}.
Let us consider a four-dimensional Lorentzian manifold $\mathcal{M}$ with a one-parameter family of metrics $g_{\mu\nu}(\lambda)$. 
We consider sufficiently small positive value of $\lambda$. Then, the metric is expanded in $\lambda$ as
\begin{align}
    g_{\mu\nu}(\lambda) = \bar{g}_{\mu\nu}+\delta g_{\mu\nu}\lambda +\frac{1}{2}\delta^2 g_{\mu\nu}\lambda^2+\mathcal{O}(\lambda^3),
\end{align}
where $\bar{g}_{\mu\nu} \coloneqq g_{\mu\nu}(0)$, and we define
\begin{align}
    \delta g_{\mu\nu} \coloneqq \left.\frac{d g_{\mu\nu}}{d\lambda}\right|_{\lambda=0},~
    \delta^2 g_{\mu\nu} \coloneqq \left.\frac{d^2 g_{\mu\nu}}{d\lambda^2}\right|_{\lambda=0}
    .
\end{align}
The parameter $\lambda$ counts the order of perturbation.

The background $\bar{g}_{\mu\nu}$ is identified with the initial stationary black hole spacetime.
More concretely, we assume that the background spacetime contains a smooth null hypersurface $\mathcal{H}$ foliated by closed future marginally outer trapped surfaces (MOTSs).
Here and throughout, ``closed'' applied to surfaces means compact without boundary.
Although we mainly have a stationary black hole as an initial spacetime in mind,
the following argument only requires the existence of such a hypersurface $\mathcal H$. A Killing horizon of a stationary black hole provides a typical example.

The perturbations $\delta g_{\mu\nu}, \delta^2 g_{\mu\nu}, \cdots$ describe how the initial black hole is modified by matter injection. Therefore, they obey the perturbed Einstein equations with source terms $\delta T_{\mu\nu}, \delta^2 T_{\mu\nu}, \cdots$,
whose restrictions to $\mathcal{H}$ are assumed to have compact support.
Before matter injection, the spacetime should coincide with the initial black hole, so $\delta g_{\mu\nu} = \delta^2 g_{\mu\nu} = 0$ there.
In addition, we assume that the null convergence condition, that is, $R_{\mu\nu}(\lambda)k^{\mu}k^{\nu} \geq 0$ for any null $k^{\mu}$, is satisfied on $\mathcal{H}$, which follows from the null energy condition through the Einstein equations.
Here and below, $R_{\mu\nu}$ and 
$C_{\mu\nu\rho\sigma}$ denote the Ricci and Weyl tensors
of $g_{\mu\nu}$, respectively.

We use the diffeomorphism freedom to choose a gauge in which $\mathcal{H}$ remains a null hypersurface with respect to $g_{\mu\nu}(\lambda)$. 
To avoid confusion with the Killing horizon of the perturbed spacetime, which we do not assume to exist, we refer to the null hypersurface $\mathcal{H}$ as the reference null hypersurface. 
More concretely, such a gauge condition can be expressed by the Gaussian null coordinates
~\cite{Penrose:1972ui, Hollands:2006rj,Hollands:2012sf},
which cover a neighborhood of $\mathcal{H}$,
as 
\begin{align}
    g_{\mu\nu}(\lambda)dx^\mu dx^\nu &= -2dv\biggl(du+\frac{u^2}{2}\alpha(\lambda)dv+u\beta_A(\lambda)dx^A\biggr) \notag\\
    & \quad \,+\gamma_{AB}(\lambda)dx^A dx^B,
\end{align}
where the null hypersurface $u = 0$ is identified with $\mathcal{H}$.
Here $x^A (A = 1,2)$ are coordinates on the two-dimensional surfaces of constant $u$ and $v$, and $\alpha(\lambda), \beta_{A}(\lambda), \gamma_{AB}(\lambda)$ are functions of $u,v,x^{A}$ which are regular at $u = 0$.

 We also introduce vector fields
 $k_{+}^{\mu} \coloneqq (\partial_v)^\mu$
 and
 $k_{-}^{\mu} \coloneqq (\partial_u)^\mu$.
 Note that $k_{\pm}^{\mu}$ are independent of $\lambda$ and the following properties hold for any $\lambda$.
 The vector field $k_{+}^{\mu}$ is an affinely parameterized null geodesic generator of $\mathcal{H}$. The vector field $k_{-}^{\mu}$ is tangent to a null geodesic congruence. These vectors are orthogonal to $u=0, v=\text{constant}$ surfaces $\mathcal{C}_{v}$, which form a foliation of $\mathcal{H}$. Therefore, the vectors $k_{+}^{\mu}$ and $k_{-}^{\mu}$ form two independent null directions normal to a surface $\mathcal{C}_{v}$.
Then, the expansions along each null congruence can be defined as 
\begin{align}
    \theta_+(\lambda) \coloneqq \gamma^{\mu}{}_{\nu}(\lambda) \nabla_\mu k_{+}^{\nu},\\
    \theta_-(\lambda) \coloneqq \gamma^{\mu}{}_{\nu}(\lambda) \nabla_\mu k^{\nu}_{-},
\end{align}
and they describe the expansion rate of the cross sectional area $\mathcal{C}_{v}$ with respect to the metric $g_{\mu\nu}(\lambda)$.
Note that the covariant derivatives appearing in the expression above are those associated with the perturbed metric $g_{\mu\nu}(\lambda)$.

Our purpose is to show the formation of a trapped surface. That is, we will show that both $\theta_{+}(\lambda)$ and $\theta_{-}(\lambda)$ are negative after matter injection. Let us expand the expansions with respect to $\lambda$,
\begin{align}
    \theta_{\pm}(\lambda)=\bar{\theta}_{\pm} +\lambda\delta\theta_{\pm} + \frac{1}{2}\lambda^2\delta^2\theta_{\pm} +\mathcal{O}(\lambda^3),
\end{align}
where $\bar{\theta}_{\pm}$ represents the expansion of $\mathcal{C}_{v}$ measured by the background metric $\bar{g}_{\mu\nu}$. Since $\mathcal{C}_{v}$ is a future MOTS with respect to $\bar{g}_{\mu\nu}$, we obtain
\begin{align}
    \bar{\theta}_+=0, \quad \bar{\theta}_-<0.  \label{theta bar}
\end{align}

For the inward expansion $\theta_{-}$, we can conclude that it remains negative $\theta_{-}(\lambda) < 0$ for sufficiently small $\lambda$, because it has a negative background value.

We then investigate $\theta_{+}(\lambda)$.
The Raychaudhuri equation for $\theta_{+}(\lambda)$ is given by
\begin{align}
    \frac{d\theta_{+}(\lambda)}{d v}=-\frac{1}{2}\theta_{+}(\lambda)^2-||\sigma_{+}(\lambda)||^2 -R_{\mu\nu}(\lambda)k_{+}^{\mu}k_{+}^{\nu}~,
\end{align}
where $||\sigma_{+}(\lambda)||^2$ is the squared norm of the shear tensor $(\sigma_+)_{\mu\nu}(\lambda)$ associated with $k_+^\mu$.
Eq.~\eqref{theta bar}, together with the background null convergence condition, implies 
\begin{align}
(\bar{\sigma}_{+})_{\mu\nu} = 0, \qquad \bar{R}_{\mu\nu}k_{+}^{\mu} k_{+}^{\nu} = 0.
\end{align}
Then, using this relation,
the Raychaudhuri equation for the first order perturbations is
\begin{align}
    \frac{d \delta \theta_+}{d v} = - \delta R_{\mu\nu}k_{+}^{\mu} k_{+}^{\nu},
\end{align}
where the right hand side is not positive under the null convergence condition.
Thus, 
if $\delta R_{\mu\nu}k_{+}^{\mu} k_{+}^{\nu} \neq 0$ at least at one point on each null generator of $\mathcal{H}$,
we conclude that, for sufficiently large $v$, $\theta_{+}(\lambda)$ becomes negative and $\mathcal{C}_v$ is a trapped surface in first order perturbations after energy injection.

In the case where $\delta R_{\mu\nu} k_{+}^{\mu} k_{+}^{\nu} = 0$ everywhere on a null generator of $\mathcal{H}$, 
we obtain $\delta \theta_{+} = 0$, and therefore we need to check the second order perturbations.
In this case, the Raychaudhuri equation for the second order perturbations is given by
\begin{align}
     \frac{d \delta^2 \theta_+}{d v}&=- 2\delta\sigma_{\mu\nu}\delta\sigma^{\mu\nu}-\delta^2 R_{\mu\nu}k_{+}^{\mu} k_{+}^{\nu}.
\end{align}
    
Again, under the null convergence condition, 
the right hand side is not positive.
In particular, it has a negative contribution when the null generic conditions~\cite{Hawking:1973uf,Wald:1984rg} are satisfied. That is, at some point on each null generator of $\mathcal{H}$, the strict null convergence condition, $\delta^2 R_{\mu\nu}k_{+}^{\mu} k_{+}^{\nu} > 0$, or $\delta ( k_{+[\mu}C_{\nu]\rho\sigma[\alpha}k_{+\beta]}k_{+}^{\rho}k_{+}^{\sigma}) \neq 0$ is satisfied.
The latter condition indicates $\delta (\sigma_{+})_{\mu\nu} \neq 0$ through the shear evolution equation.
In this case, the outward expansion $\theta_{+}$ becomes negative on a sufficiently late cross section $\mathcal{C}_v$ after the matter injection.

In summary, for sufficiently large $v$ and sufficiently small $\lambda$, expansions $\theta_{\pm}$ at the section $\mathcal{C}_{v}$ become negative up to the second order in perturbations,
\begin{align}
\label{eq:expansion_afterperturbation}
    \theta_+(\lambda) < 0, \quad \theta_-(\lambda) < 0,
\end{align}
namely, $\mathcal{C}_{v}$ becomes a trapped surface,  
if the following two conditions are both satisfied: \\
(i) the null convergence condition,
\begin{align}
 R_{\mu\nu}(\lambda) k_{+}^{\mu} k_{+}^{\nu} \geq 0,
\end{align}
is satisfied everywhere on $\mathcal{H}$,\\
(ii) the following perturbative null generic condition: at least one of the following three conditions is satisfied at some point on each generator of $\mathcal{H}$:
\begin{align}
&\delta R_{\mu\nu} k_{+}^{\mu} k_{+}^{\nu} > 0,\\
&\delta^2 R_{\mu\nu} k_{+}^{\mu} k_{+}^{\nu} > 0,\\
&\delta (k_{+[\mu}C_{\nu]\rho\sigma[\alpha}k_{+\beta]}k_{+}^{\rho}k_{+}^{\sigma}) \neq 0.
\end{align}

This gives our trapped surface criterion. 
Matter injection satisfying the above conditions produces a trapped surface $\mathcal{C}_v$. 
Then, \textit{any candidate final spacetime incompatible with the existence of this trapped surface is excluded as a possible final state.}

We emphasize that this criterion is local in the sense that it concerns only whether a candidate final spacetime can describe a neighborhood of $\mathcal{C}_v$ at sufficiently large $v$.

We close this section by noting two aspects of our trapped surface criterion that are not needed for the examples below but may be useful in broader applications.
First, the argument is not tied to a specific topology of $\mathcal{C}_{v}$ or to four dimensions.
The topology of $\mathcal C_v$ is inherited from the initial future MOTSs; in the closed case considered here, the produced trapped surface has the same topology.
The same argument for trapped surface formation also
extends straightforwardly to higher dimensions.
Second, although the setup above was phrased in Einstein gravity, the argument ultimately relies only on the corresponding properties of the Ricci tensor. 
Thus, the same reasoning applies to any metric theory of gravity, provided that the assumptions stated above in terms of the energy-momentum tensor are reformulated as the corresponding conditions on the Ricci tensor.

\section{Trapped surfaces censor superextremal final states}
\label{sec: case studies}
In this section, we show that the candidate superextremal final states in the three examples, namely Reissner–Nordstr\"om, Reissner–Nordstr\"om–de Sitter, and
Kerr-Newman spacetimes, cannot accommodate the trapped surface $\mathcal C_v$ discussed in Sec.~\ref{sec: setup}.
Our trapped surface criterion excludes such final states in the corresponding thought experiments and provides a local geometric explanation for the preservation of the weak cosmic censorship conjecture in these examples.

\subsection{Reissner-Nordstr\"om final state and its generalization}
Suppose, toward a contradiction, that the candidate final state is described near $\mathcal C_v$, for sufficiently large $v$, by a superextremal Reissner-Nordstr\"om spacetime:
\begin{align}
    ds^2&=-f(r)dt^2+f(r)^{-1}dr^2+r^2d\theta^2+r^2\sin^2\theta d\phi^2~,
\end{align}
with
\begin{align}
    f(r)&=1-\frac{2M}{r}+\frac{Q^2}{r^2},
\end{align}
and $|Q| > M$.
This spacetime has the Killing vector $(\partial_t)^{\mu}$, and importantly, it is timelike in the entire region.

The Mars-Senovilla theorem~\cite{Mars:2003ud} states that a closed trapped surface cannot be entirely contained in a region where a Killing vector is timelike.
In the present case, this timelike-Killing region is global.
Therefore, by the trapped surface criterion, a superextremal Reissner-Nordstr\"om spacetime cannot be formed from any initial black hole with a closed MOTS.
 
Let us emphasize that this argument is not tied to the explicit form of the Reissner-Nordstr\"om metric.
It applies to any final state with a global timelike Killing vector.
Examples are provided by naked-singularity solutions in the Weyl class, such as the Curzon-Chazy and Zipoy-Voorhees spacetimes~\cite{Curzon:1925, Chazy:1924, Zipoy:1966btu, Voorhees:1970ywo, Griffiths:2009dfa, Saito:2024hzc}.
Thus, for example, such static, axisymmetric symmetric naked singularity spacetimes cannot be formed from an initial Schwarzschild black hole even by injecting anisotropic matter.

\subsection{Reissner-Nordstr\"om-de Sitter final state}
Next, we consider a candidate final state described by a superextremal Reissner-Nordstr\"om-de Sitter spacetime and derive a contradiction. Unlike in the Reissner-Nordstr\"om case, the relevant Killing vector is timelike only inside the cosmological horizon. It is therefore necessary to verify that the candidate trapped surface lies inside the cosmological horizon.

To make this argument precise, we start from a slightly subextremal Reissner-Nordstr\"om-de Sitter black hole with a mass $\bar{M}$ and a charge $\bar{Q}$ and consider a transition to superextremal Reissner-Nordstr\"om-de Sitter spacetime with a mass $M(\lambda)$ and a charge $Q(\lambda)$.
To make the test nontrivial, we do not keep the initial black hole fixed away from extremality while taking $\lambda$ sufficiently small. Instead, we introduce the subextremality parameter
\begin{align}  
    \epsilon^2 \coloneqq \frac{\bar{M}- M_{\rm ext}(\bar{Q},\Lambda)}{\bar{M}},
\end{align}
and keep track of the orders of $\epsilon$ and $\lambda$ independently.
Here, $M_{\text{ext}}(Q,\Lambda)$ represents the mass of the extremal black hole with a charge $Q$ and cosmological constant $\Lambda$, given by
\begin{align}
M_{\text{ext}}(Q,\Lambda) = 
\frac{1 + 4 Q^2 \Lambda - \sqrt{1 - 4 Q^2 \Lambda}}{3 \sqrt{2 \Lambda}   \sqrt{1 - \sqrt{1 - 4 Q^2 \Lambda }}}.
\end{align}
The physically relevant regime for testing the formation of a superextremal final state is the near-extremal one, in which $\epsilon$ is comparable to the size of the perturbation.
Therefore, we focus on the limit of small $\lambda$ and $\epsilon$, with other parameters, such as $\bar{M}$ and $\Lambda$ held fixed. 

Our final state assumption allows us to express the metric around $\mathcal{C}_{v}$, for sufficiently large $v$, by the Reissner-Nordstr\"{o}m-de Sitter spacetime in the ingoing Eddington-Finkelstein coordinates as 
\begin{align}
 &g_{\mu\nu}(\lambda) dx^{\mu} dx^{\nu} \notag\\
    &= -f(\lambda; r)\,dV^2 + 2 d V d r + r^2 d\theta^2 + r^2 \sin^2\theta\, d\phi^2,
\end{align}
with
\begin{align}
    f(\lambda;r)=1-\frac{2M(\lambda)}{r}+\frac{Q^2(\lambda)}{r^2}- \frac{\Lambda}{3}r^2~.
\end{align}
In the $\lambda \to 0$ limit, the final state coincides with the initial state.
Therefore, $M(0) = \bar{M}$ and $Q(0) = \bar{Q}$ correspond to those of the initial black hole.

Let us parameterize the location of $\mathcal{C}_{v}$ in the Eddington-Finkelstein coordinates as
\begin{align}
r &= r_{\mathcal{C}_{v}}(\epsilon,\lambda; \theta, \phi),\\
V &= V_{\mathcal{C}_{v}}(\epsilon,\lambda; \theta, \phi).
\end{align}
Since the spacetime reduces to the initial black hole in the limit $\lambda \to 0$, they can be expanded as
\begin{align}
    r_{\mathcal{C}_{v}}(\epsilon, \lambda) &= \bar{r}_{\mathcal{C}_{v}}(\epsilon) +\mathcal{O}(\lambda), \\
     V_{\mathcal{C}_{v}}(\epsilon, \lambda) &= \bar{V}_{\mathcal{C}_{v}}(\epsilon) +\mathcal{O}(\lambda),
\end{align}
where $\bar{r}_{\mathcal{C}_{v}}$ and $\bar{V}_{\mathcal{C}_{v}}$ are the coordinate values of the Killing horizon in the background spacetime.

Similarly,
even in the case where the cosmological horizon is located in the neighborhood of $\mathcal{C}_v$, 
the location can be evaluated as 
\begin{align}
    r_{\text{cosmo}}(\epsilon, \lambda)  = \bar{r}_{\text{cosmo}}(\epsilon)  +\mathcal{O}(\lambda).
\end{align}
Since $\bar{r}_{\mathcal{C}_{v}} < \bar{r}_{\text{cosmo}}$ and, in the regime considered here, their separation 
is non-zero and $\mathcal{O}(\lambda^{0},\epsilon^{0})$, we can conclude that
\begin{align}
r_{\mathcal{C}_{v}}(\epsilon, \lambda)  < r_{\text{cosmo}}(\epsilon, \lambda),
\end{align}
for sufficiently small $\lambda$.
Therefore, the candidate trapped surface $\mathcal{C}_{v}$ lies inside the cosmological horizon.

Since the static Killing vector is timelike inside the cosmological horizon, we can apply the Mars-Senovilla theorem to the region containing $\mathcal C_v$.
It follows that $\mathcal C_v$ cannot be a closed trapped surface.
This contradicts the conclusion of the trapped surface criterion, and we can exclude the superextremal Reissner-Nordstr\"om-de Sitter final state in this setup.

\subsection{Kerr-Newman final state}
Finally, let us consider a superextremal Kerr-Newman spacetime as a candidate final state.
Unlike the previous two examples, Kerr-Newman spacetime contains an ergoregion, and therefore it is difficult to apply Mars-Senovilla's theorem.
Nonetheless, we can show that the candidate surface $\mathcal{C}_{v}$ is actually not a trapped surface by directly evaluating the product of the two null expansions.

As in the Reissner-Nordstr\"om-de Sitter case, we consider the setup where the initial black hole is also described by the Kerr-Newman family, with a mass $\bar{M}$, an angular momentum $\bar{J}$, and a charge $\bar{Q}$. We use the specific angular momentum $\bar{a} \coloneqq \bar{J}/\bar{M}$.
We assume $\bar{a}>0$ without loss of generality.
Then, the subextremality parameter for the initial black hole is defined by
\begin{align}
   \epsilon^2 \coloneqq \frac{\bar{M}-M_{\text{ext}}(\bar{a},\bar{Q})}{\bar{M}},
\end{align}
and we assume it is sufficiently small but comparable to the order of perturbations.
Here the mass of the extremal black hole $M_{\text{ext}}(a,Q)$ is given by
\begin{align}
    M_{\text{ext}}(a,Q) = \sqrt{a^2 + Q^2}.
\end{align}

From the assumption that the final state is the superextremal Kerr-Newman spacetime, for sufficiently large $v$, the metric near $\mathcal C_v$ can be expressed as
\begin{align}
    &g_{\mu\nu}(\lambda)dx^{\mu} dx^{\nu} \notag\\
    & = 
    - \frac{\Delta}{\Sigma} \left(
    dV - a \sin^2 \theta d \phi
    \right)^2
    + 2 dr \left( dV - a \sin^2 \theta d \phi \right) \notag\\
    &\qquad + \Sigma d \theta^2
    + \frac{\sin^2 \theta}{\Sigma} \left((r^2 + a^2) d \phi - a d V \right)^2
\end{align}
in the ingoing Eddington-Finkelstein coordinates.
Here
\begin{align}
\Sigma(\lambda) &= r^2 + a(\lambda)^2 \cos^2 \theta, \\
\Delta(\lambda) &= r^2 - 2 M(\lambda) r + a(\lambda)^2 + Q(\lambda)^2,
\end{align}
and the parameters satisfy the superextremal condition
\begin{align}
& M(\lambda) - M_{\text{ext}}(a(\lambda), Q(\lambda)) = \epsilon^2 \bar{M} + \lambda (\delta M - \delta M_{\text{ext}})\notag\\
& \qquad + 
\frac{1}{2} \lambda^2 (\delta^2 M - \delta^2 M_{\text{ext}}) + 
\mathcal{O}(\lambda^3)
 < 0, \label{superextremal condition}
\end{align}
up to the second order in perturbations,
where we introduce the perturbations of the mass of the extremal black hole as 
\begin{align}
\delta M_{\text{ext}} \coloneqq
 \left. \frac{d}{d\lambda} M_{\text{ext}}(a(\lambda),Q(\lambda)) \right|_{\lambda = 0}
= \frac{\bar{a} \delta a + \bar{Q} \delta Q}{\sqrt{\bar{a}^2 + \bar{Q}^2}},
\end{align}
and so on.

Although the relation between the Gaussian null coordinates and the ingoing Eddington-Finkelstein coordinates is unspecified, we can write the location of $\mathcal{C}_{v}$ schematically as $V = V_{\mathcal{C}_{v}}(\epsilon,\lambda; \theta, \phi)$ and $r = r_{\mathcal{C}_{v}}(\epsilon,\lambda; \theta, \phi)$.
Since the final state reduces to the initial state Kerr-Newman black hole in the limit $\lambda \to 0$, 
$V_{\mathcal{C}_{v}}$ and $r_{\mathcal{C}_{v}}$ can be expanded as 
\begin{align}
    V_{\mathcal{C}_{v}}(\epsilon,\lambda;\theta,\phi) &=\bar{V}_{\mathcal{C}_{v}}(\epsilon)+\lambda \delta V_{\mathcal{C}_{v}}(\theta,\phi)+\mathcal{O}(\lambda^2), \label{tCv Kerr} \\
    r_{\mathcal{C}_{v}}(\epsilon,\lambda;\theta,\phi)&=\bar{r}_{\mathcal{C}_{v}}(\epsilon)+\lambda \delta r_{\mathcal{C}_{v}}(\theta,\phi)+\mathcal{O}(\lambda^2)~, \label{rCv Kerr}
\end{align}
where $\bar{V}_{\mathcal{C}_{v}}(\epsilon)$ and $\bar{r}_{\mathcal{C}_{v}}(\epsilon)$ correspond to the values for the initial Kerr-Newman black hole. That is, $\bar{V}_{\mathcal{C}_{v}}(\epsilon)$ is a constant and $\bar{r}_{\mathcal{C}_{v}}(\epsilon)$ is given by
\begin{align}
\bar{r}_{\mathcal{C}_{v}}(\epsilon) &\coloneqq \bar{M} + \sqrt{\bar{M}^2 - \bar{a}^2 - \bar{Q}^2}  \notag\\
& = \bar{M} + \epsilon \sqrt{2} \bar{M} +\mathcal{O}(\epsilon^3).
\end{align}

Since the metric and the location of the surface are given, we can directly calculate the expansions, $\theta_{+}$ and $\theta_{-}$, 
along two future null directions normal to the surface given by Eqs.~\eqref{tCv Kerr} and \eqref{rCv Kerr}.
Note that although $\theta_{\pm}$ generally differ from the ones defined in Sec.~\ref{sec: setup}, the choice of the null directions is irrelevant to the trapped nature. 
Then, we can conclude that the surface $\mathcal{C}_{v}$ is not a trapped surface if 
\begin{align}
    \theta_+\theta_- < 0 
\end{align}
at a point on $\mathcal{C}_{v}$. Therefore, let us evaluate the sign of the product $\theta_{+} \theta_{-}$.

At the first order in $\lambda$ and $\epsilon$, the product of the expansions can be evaluated as 
\begin{align}
   \zeta^2 \eta(\theta) \theta_+ \theta_-&=\frac{\delta M-\delta M_{\text{ext}}}{\bar{M}}\lambda \notag\\
 &\quad + \frac{1}{ \bar{M}}\hat{L}\delta r_{\mathcal{C}_{v}} (\theta,\phi)
\lambda + \mathcal{O}(\lambda^2,\epsilon\lambda
    )~,
\end{align}
where we introduce a differential operator $\hat{L}$ by
\begin{align}
    \hat{L} &=\frac{\zeta}{\sin\theta}\frac{\partial}{\partial\theta}\left( \sin\theta \frac{\partial}{\partial \theta} \right) \notag\\
    & \quad + \frac{\eta(\theta)^2}{4 \bar{M}^4 \zeta \sin^2\theta}\frac{\partial^2}{\partial \phi^2} + \frac{\bar{a}}{\bar{M}} \frac{\partial}{\partial \phi}~.
\end{align}
We have introduced a constant $\zeta$ and a positive function $\eta$ by
\begin{align}
   \zeta &\coloneqq \frac{\bar{M}^2 + \bar{a}^2}{2 \bar{M}^2}, \\
    \eta(\theta) &\coloneqq  \bar{M}^2 + \bar{a}^2 \cos^2 \theta.
\end{align}
The first term is non-positive, $\delta M - \delta M_{\text{ext}} \leq 0$, which follows from Eq.~\eqref{superextremal condition} omitting $\mathcal{O}(\epsilon^2, \lambda^2)$.
 Let $d\Omega=\sin\theta d\theta d\phi$ be the standard measure on $S^2$. One can show that the integration of
$\hat{L}\delta r_{\mathcal{C}_{v}}(\theta,\phi)$ over the sphere vanishes:
\begin{align}
    \int_{S^2} \hat{L} \delta r_{\mathcal{C}_{v}} (\theta,\phi) d\Omega=    0~.
\end{align}
Therefore, \(\hat{L}\delta r_{\mathcal{C}_{v}}(\theta,\phi)\) cannot be strictly positive everywhere on the sphere.
In other words, there are two possibilities: (i) \(\hat{L}\delta r_{\mathcal{C}_{v}}\) becomes negative at some points on \(\mathcal{C}_{v}\), or (ii) \(\hat{L}\delta r_{\mathcal{C}_{v}}=0\) everywhere on \(\mathcal{C}_{v}\).
For the case (i), the surface \(\mathcal{C}_{v}\) is not a trapped surface already at the first order in \(\lambda\) and \(\epsilon\).
For the case (ii), where \(\hat{L}\delta r_{\mathcal{C}_{v}}=0\) everywhere, the first-order analysis is still sufficient to conclude the absence of a trapped surface when \(\delta M-\delta M_{\text{ext}}<0\).
Therefore, the only case not resolved at first order is
\[
\delta M-\delta M_{\text{ext}}=0,
\qquad
\hat{L}\delta r_{\mathcal{C}_{v}}=0
\quad \text{everywhere on } \mathcal{C}_{v}.
\]
For this case, the first-order contribution to the expansion vanishes identically, so we need to proceed to the second-order analysis.

Let us investigate the case (ii) with $\delta M = \delta M_{\text{ext}}$ in more detail.
From the expression
\begin{align}
    &\int_{S^2} \delta r_{\mathcal{C}_{v}} \hat{L} \delta r_{\mathcal{C}_{v}}(\theta,\phi) d\Omega \nonumber\\
    &~~=-\zeta \int_{S^2} (\partial_{\theta} \delta r_{\mathcal{C}_{v}})^2 d\Omega \notag\\
    & \qquad - \int_{S^2}\frac{\eta^2(\theta)}{4 \bar{M}^4 \zeta \sin^2  \theta}(\partial_{\phi} \delta r_{\mathcal{C}_{v}})^2d\Omega~,
\end{align}
one can see that
$\hat{L} \delta r_{\mathcal{C}_{v}} = 0$ everywhere on $\mathcal{C}_{v}$ if and only if
\begin{align}
    \delta r_{\mathcal{C}_{v}}(\theta,\phi)= \text{const.}
\end{align}
Using this condition, the product of the expansions up to the second order in $\lambda$ and $\epsilon$ can be evaluated as 
\begin{align}
    & \zeta^2  \eta(\theta) \theta_+(\lambda)\theta_-(\lambda) = \left(\frac{M(\lambda)-M_{\text{ext}}(a(\lambda),Q(\lambda))}{\bar{M}}\right) \notag\\
    & \qquad 
     -\left(\epsilon -\frac{\lambda  (\delta M - \delta r_{\mathcal{C}_{v}})}{\sqrt{2} \bar{M}}\right)^2
    + \frac{1}{2 \bar{M}} \hat{L} \delta^2 r_{\mathcal{C}_{v}} \lambda^2
     \notag\\
    & \qquad +\mathcal{O}(\lambda^3,
    \lambda^2\epsilon,\lambda\epsilon^2),
\end{align}
The first term on the right-hand side is negative for a superextremal final state, $M(\lambda) < M_{\text{ext}}(a(\lambda),Q(\lambda))$ up to second order.
The second term is non-positive. By the same argument used at first order, the third term is
either negative somewhere on $\mathcal C_v$ or vanishes identically. 
We can therefore conclude that
\begin{align}
    \theta_+(\lambda)\theta_-(\lambda)  = (\text{negative term}) + \mathcal{O}(\lambda^3, \lambda^2 \epsilon, \lambda \epsilon^2),
\end{align}
and $\mathcal{C}_{v}$ is not a trapped surface for sufficiently small $\lambda$ and $\epsilon$.

Therefore, by our trapped surface criterion, we can conclude that the superextremal Kerr-Newman spacetime is excluded as a final state of the overcharging/overspinning thought experiment.

\section{Summary and Discussion}
\label{sec: summary}
We proposed a local geometric criterion for black hole overcharging/overspinning thought experiments. 
We showed that matter injection satisfying the null convergence and generic conditions turns a cross section of the initial Killing horizon into a trapped surface. 
Our criterion states that \textit{any candidate final spacetime unable to accommodate this surface is ruled out}.
Thus, in a censorship test, the relevant question is whether a proposed naked-singularity spacetime can accommodate the trapped surface.

Applying this criterion to Reissner-Nordstr\"om, Reissner-Nordstr\"om-de Sitter, and Kerr-Newman spacetimes, we showed that it rules out the candidate superextremal final states. 
These results recover the conclusions of the Sorce-Wald formalism from a local geometric viewpoint.
In addition, our criterion can exclude the formation of any spacetime with a globally timelike Killing vector. For example, the formation of the naked singularity in the Weyl class can be excluded by our criterion.

Since the argument for the trapped surface formation is not tied to four dimensions or to any particular topology, it suggests possible applications to censorship tests for higher-dimensional black objects with non-spherical horizon topology, such as black rings and black lenses~\cite{Emparan:2001wn,Izumi:2007qx,Kunduri:2014kja,Tomizawa:2016kjh,Emparan:2008eg,Hollands:2012xy}.
The relevant question is then whether the candidate final spacetime can accommodate a trapped codimension-two surface with this inherited topology.

The main advantage of our criterion is that it does not require asymptotically defined conserved charges. 
This feature has two possible implications. 
First, it can be useful even in asymptotically flat spacetimes when candidate final states with naked singularities are not characterized by a simple extremality bound written in terms of conserved charges. 
The static Weyl class above provides such examples. 
Second, the criterion may be applicable to spacetimes with nonstandard asymptotic structures, such as black holes immersed in external magnetic fields~\cite{Wald:1974np,Ernst:1976mzr,Ernst:1976bsr,aliev1988energetics,Aliev:1989wz,Gibbons:2013yq,Podolsky:2025tle,Ovcharenko:2025cpm}, where the definition or interpretation of asymptotic charges is less direct.
Indeed, the approach adopted for the Kerr-Newman spacetime does not require any specific geometric structure, such as the existence of a timelike Killing vector. 
Therefore, the method is, in principle, applicable to such broader classes of spacetimes.

In Einstein gravity, the null convergence condition is equivalent to the null energy condition. While our criterion is directly applicable under the null convergence condition even in theories beyond Einstein gravity, it is worth discussing whether we can extend our formulation under a physically reasonable condition expected in each theory, such as monotonicity of an appropriate entropy~\cite{Bousso:2015mna,Wall:2015raa,Bhattacharyya:2021jhr,Hollands:2022fkn,Davies:2023qaa,Hollands:2024vbe}.
In particular, \emph{entropic trapped surfaces}~\cite{Furugori:2025pmn} provide an appropriate generalization of trapped surfaces in each theory, suggesting a possible extension of our trapped surface criterion beyond Einstein gravity.

The Kerr-Newman example also highlights a mathematical question raised by our criterion: whether superextremal spacetimes can contain closed trapped surfaces. 
While Mars-Senovilla's theorem~\cite{Mars:2003ud} gives a strong obstruction for static spacetimes with a globally timelike Killing vector, no analogous general result seems to be available for rotating spacetimes with an ergoregion. 
A general obstruction to closed trapped surfaces in such spacetimes would significantly broaden the applicability of our approach.

Overall, our results suggest that the protection of weak cosmic censorship in overcharging/overspinning processes can be understood as a local geometric obstruction: matter injection creates a closed trapped surface, and any candidate final state must be able to accommodate it. This viewpoint separates the censorship mechanism from asymptotic charge conservation and may provide a local geometric basis for testing weak cosmic censorship in more general dynamical settings.

\begin{acknowledgments}
We thank Jos\'{e} M. M. Senovilla for valuable discussions about possible applications of the Mars-Senovilla theorem to superextremal spacetimes. This work was supported by JSPS KAKENHI Grant No.~JP22H01217~(H.F.) and No.~JP26K07084~(D.Y.). K.Y. is supported by JST SPRING, Grant Number JPMJSP2108 and by a research encouragement grant from the Iwanami Fūjukai.
\end{acknowledgments}

\bibliography{ref}
\end{document}